\documentclass[aip,graphicx]{revtex4-1}
\usepackage{graphicx}
\draft 
\begin{document}


\title{Status of the GINGER project}

\author{Carlo Altucci$^{1,2}$, Francesco Bajardi$^{1,3}$, Andrea Basti$^{4,5}$, Nicol\`o Beverini$^4$, Giorgio Carelli$^{4, 5}$, Salvatore Capozziello$^{1,6}$, Simone Castellano$^{5,7}$, Donatella Ciampini$^{4, 5}$, Fabrizio Davì$^8$, Francesco dell'Isola$^9$, Gaetano De Luca$^{10}$, Roberto Devoti$^{10}$, 
Giuseppe Di Somma$^{4,5}$, Angela D.V. Di Virgilio$^5$,  Francesco Fuso$^{4,5}$,
Ivan Giorgio$^9$, Aladino Govoni$^{10}$,  Enrico Maccioni$^{4,5}$, Paolo Marsili$^{4,5}$, Antonello Ortolan$^{11}$, Alberto Porzio$^{1,12}$, Matteo Luca Ruggiero$^{13}$, Raffaele Velotta$^{1,3}$}
\address{
$^1$ Istituto Nazionale di Fisica Nucleare, Sezione di Napoli, Via Cintia 26, 80126 Napoli, Italy\\
$^2$ Department of Advanced Biomedical Sciences, Universit\`a  degli Studi di
Napoli Federico II, Via Cintia 26, 80126 Napoli, Italy.\\
$^3$ Scuola Superiore Meridionale, Largo San Marcellino 10, 80138 Napoli, Italy\\
$^4$ Department of Physics, University of Pisa, Largo B. Pontecorvo 3, 56127 Pisa, Italy.\\
$^5$ Istituto Nazionale di Fisica Nucleare, Sezione di Pisa, Largo B. Pontecorvo 3, 56127 Pisa, Italy\\
$^6$ Department of Physics, Universit\`a  degli Studi di
Napoli, Federico II, Via Cintia 26, 80126 Napoli, Italy.\\
$^7$ Gran Sasso Science Iinstitute,  Viale Francesco Crispi, 7, 67100 L'Aquila, Italy.\\
$^8$ Universit\`a Politecnica delle Marche Via Brecce Bianche 12,  60131 Ancona, Italy.\\
$^9$ DICEAA and Memocs Universit\`a dell'Aquila, Piazzale E. Pontieri 1 - Monteluco di Roio, 67100 L’AQUILA, Italy.\\
$^{10}$ Istituto Nazionale di Geofisica e Vulcanologia, via delle Vigna Murata 605, Rome, Italy\\
$^{11}$ Istituto Nazionale di Fisica Nucleare, Legnaro INFN National Laboratory, Viale dell’Universit\`a 2, 35020 Legnaro (PD), Italy.\\
$^{12}$  DICEM, Universita di Cassino e Lazio Meridionale, 03043, Cassino, Italy\\
$^{13}$ Department of Mathematics, University of Turin, Via Verdi 8 - 10124 Torino, Italy}
\date{\today}
\begin{abstract}
Large frame Ring laser gyroscopes, based on the Sagnac effect, are top sensitivity instrumentation to measure  angular velocity with respect to the fixed stars. GINGER (Gyroscopes IN GEneral Relativity) project foresees the construction of an array of three large dimension ring laser gyroscopes, rigidly connected to the Earth. GINGER has the potentiality to measure  general relativity effects and Lorentz Violation in the gravity sector, once a sensitivity of $10^{-9}$, or better, of the Earth rotation rate is obtained. Being attached to the Earth crust, the array will also provide useful data for geophysical investigation. For this purpose, it is at present under construction as part of the multi-components observatory called Underground Geophysics at Gran Sasso (UGSS). Sensitivity is the key point to determine the relevance of this instrument for fundamental science. The most recent progress in the sensitivity measurement, obtained on a ring laser prototype called GINGERINO, indicates that GINGER should reach the level of 1 part in $10^{11}$ of the Earth rotation rate.
\end{abstract}
\maketitle 
\section*{}

When the frame supporting an optical cavity rotates, the  interferometer is sensitive to the cavity rotation rate, an effect usually called Sagnac effect.  Accordingly, Sagnac interferometers are a specific class of interferometers, commonly used to measure inertial  angular velocity. The Sagnac effect largely dominates all other effects and is  generally exploited in devices for inertial navigation.
The   active large frame Sagnac interferometer, usually referred to as  ``Ring Laser Gyro'' (RLG), is by far the most sensitive instrument in this family and has shown long time continuous operation capabilities. A sensitivity of the order of prad/s with long term continuous operation and large dynamic range have been extensively demonstrated\cite{uno, due}.
An RLG is built as a closed path interferometer, usually defined by 4 mirrors located at the vertices of a square, and two counter propagating laser beams are excited inside the optical cavity. 
The interference of the beams transmitted by each mirror brings information on the non reciprocal effects experienced by the two counter-propagating beams due to the geometry or the laser dynamic. Since the interferometer has two equal paths, the differences due to these non reciprocity effects are extremely small. 
However, there are other non reciprocal effects related to the space time structure or to fundamental asymmetries, and so RLGs are suitable for fundamental physics investigations.

 In general a large frame RLG has a square optical cavity, above 3-4~m sides, and it operates rigidly attached to the ground. An RLG senses the component of the angular velocity vector $\vec{\Omega}$ along the axis of the closed polygonal cavity, defined by the area vector. The relationship between the Sagnac frequency $\omega_s$ and the angular rotation rate $\Omega$ reads: 

\begin{equation}
\omega_s =4\frac{A}{\lambda L} \Omega \cos{\theta} \ , \\
\end{equation}

where $A$ is the area enclosed by the optical path, $L$ is its perimeter, $\lambda$ is the wavelength of the light, and $\theta$ is the angle between the area vector  and  $\vec{\Omega}$. 

The GINGER project foresees 3 RLGs attached to the Earth crust; however, in this first stage  only two of them will be built.  The main objective of the instrument is to reconstruct the total angular velocity vector $\vec{\Omega}$, which contains, beside the kinematic term $\vec{\Omega}_\oplus$,  the contributions due to gravity (e.g. DeSitter and Lense-Thirring effects of General Relativity), to the  Lorentz violation (if any) in the framework of Standard Model Extended (SME). The kinematic local and global contributions are derived from geophysics and geodesy, e.g. the sub-daily component of the Length of Day (LOD). These kinematic terms are continuously monitored by the International Earth Rotation System (IERS) with very high accuracy, and so gravitational theories can be tested by comparing the independent measurements of RLGs and IERS. 
    
The effectiveness of GINGER for fundamental physics investigations depends on its sensitivity, which quite often is expressed as relative precision in the measurement of the Earth angular rotation rate. It can be said that an accuracy of 1 part in $10^{9}$ is the target to be meaningful for fundamental physics, at present 1 part in $10^{11}$ seems  feasible for GINGER\cite{PRD,Tartaglia,Capozziello, quattro}.
The experimental set up plays a big role, since the response depends on the geometry and it is necessary to avoid spurious rotation of the apparatus induced by environmental disturbances. The first working RLGs were based on monolithic structures made of very low thermal expansion ceramic materials, and most of the small size gyroscopes are monolithic. However, for large very high performance apparatus,  this choice is quite challenging in terms of cost and space, and it poses severe limitations in its use in  an array. A heterolytic (HL) cavity is composed of different mechanical components, whose relative orientation can be modified. Therefore, rigidity and geometrical stability must be ensured bu using, if necessary,  PZT-driven active stabilisation servos. Two HL prototypes have been built and extensively studied by our group: GINGERINO \cite{GING1}, placed in the Gran Sasso underground INFN laboratories, and GP2 \cite{GP2}, located in Pisa INFN laboratories. Our experimental work has indicated that GINGER can be realized in a HL structure, and that an underground location is particularly recommended, since it takes advantage of the natural thermal stability and of reduced environmental noise. Unattended continuous operation for months, a typical sub-prad/s sensitivity in 1 second of measurement time, large bandwidth, fast response, in principle as fast as milliseconds, have been proven in the experiments carried out so far.\\
At present, beside our RLGs, several other high sensitivity instruments are in operation: in Germany, G \cite{G1} of the geodetic observatory of Wettzell, and ROMY, an array of four triangular rings at the geophysical observatory of Bavaria\cite{ROMY},  ER1 at the University of Canterbury,\cite{ER1} in New Zealand, and HUST-1 \cite{HUST} a passive gyroscope built at Huazhong University of Science and Technology in Wuhan  (China) as part of the TianQin project. Collaboration with all these  groups is already active.\\
In order to fully define  the angular rotation vector, at least 3 RLGs are required, and redundancy would be  welcome in this kind of high sensitivity apparatus.  The GINGER project foresees the construction of an array of 3 equal RLGs\cite{FeasStudy}.
The first one, called RLX,  oriented at the maximum Sagnac frequency, with its axis parallel to Earth rotational axis, will provide the absolute value of the angular velocity; the second, RLH, with the axis along the local vertical, and the third one, RLO, will be oriented outside the meridian plane, to complete the vector  $\vec{\Omega}$ determination. The sensitivity is such that very small variations in the orientation of RLH and RLO will affect the measurement; RLX allows a precise determination of the orientation of the other two gyros with respect to the rotation axis \cite{FeasStudy}.
\\ 
Sensitivity is a key point. 
The limiting noise is determined by the shot noise of the apparatus, which is a function of the cavity losses. By using the parameters of our prototype GINGERINO, the classical shot noise model\cite{uno,chow} estimates 50 prad/s in 1 second measurement. 
Recently, we have been able to evaluate directly the limiting noise of the GINGERINO prototype, demonstrating, thanks to a new detection scheme, that  the limiting noise floor is in the prad/s Hz$^{-1/2}$ range (in the frequency range  $< 0.1$ Hz), more than a factor 10 below the expected one\cite{SN2023}.  
This experimental result is clearly not compatible with the conventional shot noise evaluation, suggesting that a complete quantum model of the system is necessary. 
Indeed, the conventional model does not take into account couplings among the two counter-propagating beams.
In a forthcoming study we want to develop a quantum model that, tracing back from the detector scheme, accounts for all the complex interdependent dynamics of the counter--propagating beams  with the laser medium and the mirrors.
Anyway, this experimental noise level limit indicates that a realistic final sensitivity target of GINGER should be around 1 part in $10^{11}$ of the Earth rotation rate.
\\
By the end of 2023 we will define the executive design of GINGER and in 2024 we will start to mount the apparatus inside the Gran Sasso laboratories. 
The whole experimental set up has been developed based on the experience acquired on the GINGERINO apparatus and details of the experimental layout can be found in the literature \cite{MEMOC}.
In the first part of the project we plan to build only two RLGs, RLX and RLH,
with a planned perimeter of each square optical cavity of 16~m.  
Figure 1 shows a pictorial view of GINGER located inside Node B of the Gran Sasso laboratory. We expect that construction will take 18 months, so to have the array in operation between the end of 2025 and the beginning of 2026.
\\
GINGER is intended as an interdisciplinary project, with significant expected results not only in the field of fundamental physics, but also in geodesy and in geophysics. It should give  information complementary with the GNNS and VLBI networks about the Length Of the Day and the Earth polar motion.  For geophysical applications, it will provide the rotational seismic information to the multi-components geophysical observatory {\it Underground Geophysics at Gran Sasso} (UGSS).   
\\
The construction of GINGER is co-funded by INFN and INGV.
\begin{figure}
\centering
 \includegraphics[scale=0.6]{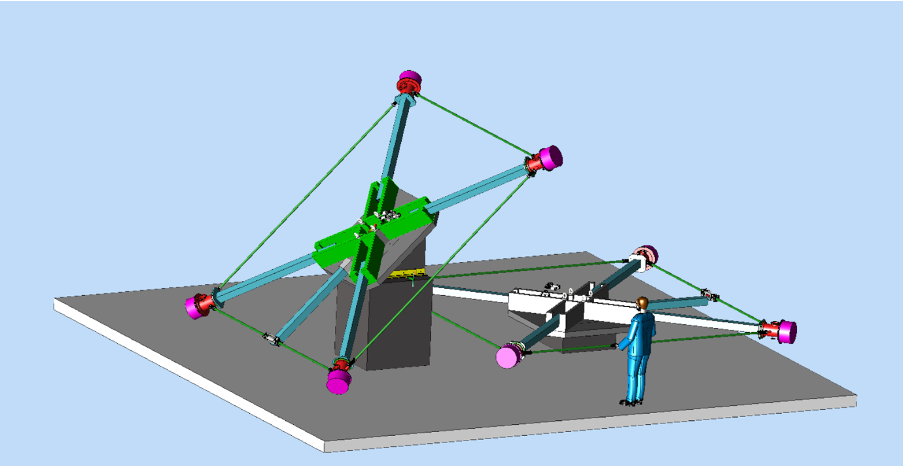}
 \caption{Pictorial view of GINGER, the two RLGs are visible.}
 \label{GINGER}
\end{figure}


%
%

%



\end{document}